\documentclass[aps,prl,twocolumn,showpacs,preprintnumbers,amsmath,amssymb,revsymb]{revtex4}
\usepackage{graphicx}
\usepackage{color} 
\usepackage{dcolumn}
\usepackage{bm}
\usepackage{braket}
\usepackage{setspace}
\usepackage{sidecap}

\definecolor{gray}{rgb}{0.5,0.5,0.5}

\def\braket#1{\mathinner{\langle{#1}\rangle}}

\providecommand{\h}{\hbar}
\providecommand{\p}{\partial}

\newcommand{\dd}{\textrm{d}} 
\newcommand{\kb}{k_{\textsc{b}}^{}}
\newcommand{\om}{\omega}

\begin{document}

\title{Interaction induced edge channel equilibration }

\preprint{}

\author{Anders~Mathias~Lunde}\email[Corresponding author: ]{mathias.lunde@unige.ch}
\author{Simon~E.~Nigg}
\author{Markus~B\"uttiker}
\affiliation{D\'epartement de Physique Th\'eorique, Universit\'e de
  Gen\`eve, CH-1211 Gen\`eve 4, Switzerland}
\date{\today}

\begin{abstract}
The electronic distribution functions of two Coulomb coupled chiral
edge states forming a quasi-1D system with broken translation
invariance are found using the equation of motion approach. We find
that relaxation and thereby energy exchange between
the two edge states is determined by the shot noise of the edge states
generated at a quantum
point contact (QPC). In close vicinity to the QPC, we derive analytic expressions for the distribution
functions. We further give an iterative procedure with which
we can compute numerically the distribution functions arbitrarily far
away from the QPC. Our results are compared with
recent experiments of Le Sueur et al..
\end{abstract}

\maketitle

\begin{figure}[hb]
\includegraphics[width=0.45\textwidth]{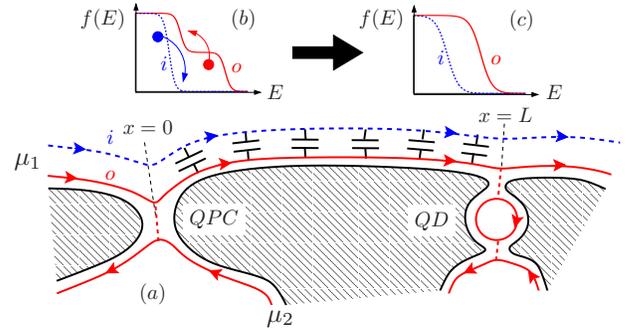}
\caption{(color online). (a) The experimental setup to measure the electronic distribution
function of an ES. The full (red) curve represents the
measured 
outer ES while the dashed (blue) curve represents a co-propagating
inner ES. The two ES's exchange energy via Coulomb interaction between $x=0$ and $x=L$. The initial distribution
functions (b) relax, via energy-conserving particle-hole
excitation processes, toward Fermi functions (c).\label{fig:geometry}}
\end{figure}

Two decades ago, edge states (ES's)~\cite{Halperin:82} were demonstrated to
be a physical reality by creating a non-equilibrium
population~\cite{Buttiker:88a} through selective injection and
detection of carriers in different states along the same
edge~\cite{Komiyama:89,Wees:89,Alphenaar:90}. Experiments revealed
that the inter-edge carrier scattering could be strongly suppressed
\cite{Komiyama:89,Wees:89,Alphenaar:90} over distances of  $80\,{\rm
  \mu m}$. Now in a series of novel experiments the group of
Pierre~\cite{Altimiras:09a,Sueur:09} has investigated the
non-equilibrium distribution function in an ES as it evolves along a channel away from a QPC at which it is initially created. The
experiments are carried out in a high mobility two-dimensional electron gas at a filling
factor $\nu=2$ such that there is an outer (spin up) non-equilibrium ES and an
inner (spin down) equilibrium ES. The distribution function is measured with the
help of a quantum dot (QD) sufficiently small to provide transmission
only through a single resonant level, see fig.~\ref{fig:geometry}. The
QD serves as an energy spectrometer and permits the reconstruction of
the distribution function in the outer ES.

The experiments
reveal two surprising features:  First, the initial non-equilibrium
distribution created at the QPC and calculated form non-interacting
scattering theory differs only weakly from the measured one over
distances of close to one micrometer \cite{Altimiras:09a,Sueur:09}. At
large distances from the QPC, due to the Coulomb interaction between
carriers in the two ES's, the distribution function evolves
into an equilibrium distribution function at an effective
electrochemical potential and temperature. The outer non-equilibrium
ES transfers part of its energy to the inner ES. The
two ES's equilibrate towards the same equilibrium
  distribution with the same temperature (but still at different
electrochemical potentials due to lack of particle exchange between
the two ES's). The second surprise of the experiments is the
fact that the temperature of the distribution
functions at large distance in the two ES's is {\it lower} than dictated by
equilibrium thermodynamic arguments \cite{Sueur:09}. The first surprise
shows that relaxation due to inter-ES interaction is weak. The second surprise implies that equilibration
occurs not only between the inner and outer ES's but that there
must be an additional equilibration mechanism which cools the two ES's below what would be expected from inter-ES
  coupling alone. We propose that additional excitations in the
bulk~\cite{Granger:09}, which couple predominantly to the inner ES, have to be considered
to understand this effect. Although the nature of these excitations
remains unclear, the experimental findings of~\cite{Sueur:09} are consistent with this
hypothesis. For example it is found that when the inner ES is
forced to form a short closed loop, then relaxation in the outer ES is strongly suppressed.

The physics of ES's is often discussed within the framework of
bosonization theory, where the elementary excitations have
bosonic character and are of collective nature~\cite{MachZehnder,Degiovanni:09}. In contrast, we take the weak
equilibration seen at distances of less than a
micrometer as the starting point of a discussion which treats inter-ES interaction perturbatively~\cite{comment-weak-int}. The interaction is described in
terms of two-body collisions. We use the equation of motion approach
for second quantized operators to derive an evolution equation for the
distribution functions which resembles a Boltzmann collision term with
the added complication that there are two different initial
distributions (one for each ES). Alternatively the Coulomb
matrix elements which appear in this theory can be taken from an RPA
theory~\cite{RPA} in which the electron densities in each channel fluctuate and
interact through an (effective) capacitance. To treat equilibration at
longer distances we iterate numerically the solution for short
distances. At large distances the distribution functions approach
their equilibrium form dictated by entropy
maximization. 

We describe the ES's in terms of scattering states
$\chi_{\alpha E}(x)$ with energy $E$ and label $\alpha=o,i$ ($i:$
inner, $o:$ outer). The inter-ES interaction is given by
\begin{align}
H_{\textrm{int}}
&=\frac{1}{2} \sum_\alpha \int \dd E\,\dd E'\, U_\alpha(E',E)\,
a^\dag_{\alpha E'} a_{\alpha E}^{},
\end{align}
where $a^\dag_{\alpha E}$ ($a_{\alpha E}^{}$) is the creation (annihilation) operator for the scattering state $\chi_{\alpha E}$ and $U_{\alpha}(E',E)$ is the potential operator for scattering a particle from $E$ to $E'$ in the ES $\alpha$ at the expense of a particle scattering in the opposite ES $\bar{\alpha}$. Explicitly 
$U_{\alpha}(E_{1'},E_1,t)=\int \dd E_2 \dd E_{2'} V^{\alpha\bar{\alpha}}_{E_{1'} E_{2'},E_1 E_2} a^\dag_{\bar{\alpha} E_{2'}}(t)a_{\bar{\alpha} E_2}^{}(t)$
in the Heisenberg picture 
and  $V^{\alpha\bar{\alpha}}_{E_{1'} E_{2'},E_1 E_2}$ is the inter-ES electron-electron interaction matrix element for the scattering
process $(\alpha E_1,\bar{\alpha} E_{2})\rightarrow(\alpha
E_{1'},\bar{\alpha} E_{2'})$. Using the Heisenberg equation of motion
$i\h \p_ta_{\alpha E}^{}(t)=[a_{\alpha E}^{}(t),H]$, the electronic
distribution function $f_\alpha(E)$ in ES $\alpha$ can be
found by evaluating $\langle a^\dag_{\alpha E}(t) a_{\alpha
  E'}^{}(t)\rangle=\delta(E-E')f_\alpha(E)$. The non-interacting
distributions are $f_i^{0}(E)=f_{\mu_i}^0(E)$ and
$f_o^{0}(E)=\mathcal{R}f_{\mu_1}^0(E)+\mathcal{T}f_{\mu_2}^0(E)$,
where $f^0_{\mu}\equiv\{1+\exp[(E-\mu)/\kb T]\}^{-1}$ and $\mathcal{T}$
($\mathcal{R}$) is the transmission (reflection) probability of the
QPC, see Fig.~\ref{fig:geometry}.  The chemical potential of the inner
ES $\mu_i$ can experimentally be tuned independently of
$\mu_1$ and $\mu_2$ by using an additional QPC (not shown in
Fig.~\ref{fig:geometry}). To second order in the interaction matrix
element the distribution is $f_\alpha^{(2)}=f_\alpha^{0}+\delta f^{(2)}_\alpha$, where (see Supplementary Material for details) 
\begin{align}
&\delta f^{(2)}_\alpha(E)= \nonumber \\
&2\pi \!\!\int_{-\infty}^\infty\!\!\!\!\!\dd \om\!
\bigg[f_\alpha^0(E+\h\om)[1-f_\alpha^0(E)]
S^a_{\delta U_{\alpha}\delta U_{\alpha}}(E,E+\h\om,\om)
\nonumber\\
&-f_\alpha^0(E)[1-f_\alpha^0(E+\h\om)]
S^e_{\delta U_{\alpha}\delta U_{\alpha}}(E+\h\om,E,\om) \bigg].
\label{eq:df-inter-second-order}
\end{align}
The first term contains the absorption potential fluctuation spectrum~\cite{Aguado:00}
$S^{a}_{\delta U_{\alpha}\delta U_{\alpha}}(E,E',\om)$ describing an
absorption of energy $\h\om$ by the ES $\bar{\alpha}$ while
the ES $\alpha$ goes from energy $E'$ to $E$. Likewise the
second term with the emission fluctuation spectrum $S^e_{\delta
  U_{\alpha}\delta U_{\alpha}}$ describes the emission of energy
$\h\om$ from the ES $\bar{\alpha}$ to the ES $\alpha$, which
consequently leads to the transition $E\rightarrow E+\h\om$ in
$\alpha$. The fluctuation spectra are to lowest order in the
interaction and defined by
%
$2\pi \delta(\om+\om')S^a_{\delta U_{\alpha}\delta U_{\alpha}}(E',E,\om)\equiv
\langle \delta U_{\alpha}(E,E',\om)^{(1)} \delta U_{\alpha}(E',E,\om')^{(1)}\rangle$,
where $\delta U_{\alpha}^{(1)}\equiv
U_{\alpha}^{(1)}-\langle U_{\alpha}^{(1)}\rangle$
is the Fourier transformed operator for the deviation from the average
potential to first order in the interaction. The emission
spectrum is found by interchanging the two $\delta U_\alpha$ in the
absorption spectrum or equivalently by changing the sign of $\omega$.
Explicitly, the spectra are found to be
\begin{subequations}
\label{eq:absob-emis-spectra}
\begin{align}
&S^a_{\delta U_{\alpha}\delta U_{\alpha}}(E',E,\om)=\nonumber \\
&h \!\int \!\!\dd E'' |V^{\alpha\bar{\alpha}}_{E'E''+\h\om,EE''}|^2
f_{\bar{\alpha}}^0(E'')
[1-f_{\bar{\alpha}}^0(E''+\h\om)],\\
&S^e_{\delta U_{\alpha}\delta U_{\alpha}}(E',E,\om)=\nonumber\\
& h \!\int \!\!\dd E'' |V^{\alpha\bar{\alpha}}_{E'E'',EE''+\h\om}|^2
f_{\bar{\alpha}}^0(E''+\h\om)
[1-f_{\bar{\alpha}}^0(E'')],
\end{align}
\end{subequations}
where the interpretation in terms of emission and absorption spectra
is clear. By inserting these into Eq.~(\ref{eq:df-inter-second-order})
the similarity with the collision integral in the Boltzmann equation
becomes evident.


Next we wish to calculate $\delta f^{(2)}_\alpha(E)$. To this end, the
inter-ES scattering process $(\alpha E_1,\bar{\alpha}
E_{2})\rightarrow(\alpha E_{1'},\bar{\alpha} E_{2'})$ needs to be
considered. If the ES's are perfectly translation invariant,
then energy \emph{and} momentum conservation together reduce the
available one dimensional phase space enormously compared
to higher dimensions \cite{Lunde-Flensberg-Glazman}. 
This leads us to consider the more
realistic non-translation invariant case caused by the fact that
the ES's follow the equipotential lines created by the sample
edges and the impurity potential. Including this non-translation
invariant ES physics leads to the presence of non-momentum
conserving scattering processes increasing the phase space
substantially \cite{Rech:08,Lunde:09}. The broken translation
invariance is included into the model of the inter-ES interaction
matrix element $|V^{\alpha\bar{\alpha}}_{E_{1'}E_{2'},E_1 E_2}|^2$. To
avoid modeling a specific geometry we perform a statistical average
over the geometry of the ES's and thereby introduce the
momentum breaking correlation length $\ell_p$, which is smaller than
the size of the region of relaxation $L$. For simplicity, an effective
interaction of the form $V(x,x')=\delta(x-x')g(x)$ is used and it is
assumed that the deviation of $g(x)$ from some mean value $g_0$ is
Gaussian distributed,
i.e.~$\overline{(g(x)-g_0)(g(x')-g_0)}=A/(\sqrt{2\pi}\ell_p)\exp\big[-(x-x')^2/(2\ell_p^2)\big]$
where $A/(\sqrt{2\pi}\ell_p)$ is the maximal deviation and
$\overline{\cdots}$ denotes the geometrical averaging. This yields an
interaction with
a momentum conserving and a momentum breaking part. The latter is (see Supplementary Material for details) 
\begin{align}
\overline{|V^{\alpha\bar{\alpha}}_{E_{1'}E_{2'},E_1 E_2}|^2}_{\Delta k\neq 0}=
\frac{AL}{h^4v_\alpha^2v_{\bar{\alpha}}^2}  \exp\left[-(\Delta k\ell_p)^2/2\right], \label{eq:matrix-element-non-momentum-conserving}
\end{align}
where $\Delta k=(E_1-E_{1'})/(\h v_\alpha)+(E_2-E_{2'})/(\h
v_{\bar{\alpha}})$, using linear dispersion relations with different
velocities $v_\alpha$ for the two ES's. 
%
Note that for linear dispersions with \emph{different} velocities
there is no phase space for scattering in the momentum conserving
limit, $\Delta k=0$, but in the very special (almost pathological)
case $v_\alpha=v_{\bar{\alpha}}$, momentum and energy conservation
are equivalent leading to plenty of phase space. The specific model for
the interaction and the matrix element is not of great importance as
long as it includes the physics leading to non-momentum conserving
processes, which in turn introduces a new length scale $\ell_p$. 
%
%

\begin{figure}
\includegraphics[width=\linewidth]{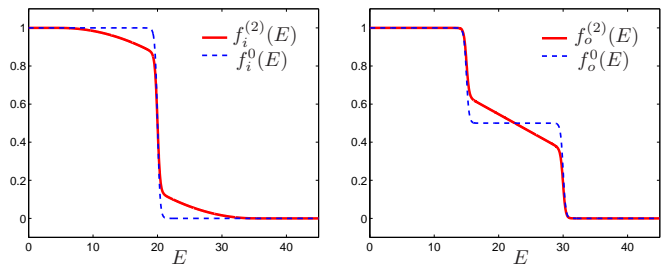}
\caption{(color online). Analytically
  calculated inner
  (left) and outer (right) ES distribution functions in the
  regime $\kb T,\lvert\mu_2-\mu_1\rvert\ll\lvert\Delta E\rvert$. The parameters are (energies in ${\rm \mu eV}$): $\mathcal{T}=0.5$, $\mu_1=15$,
  $\mu_2=30$, $\mu_i=20$, $\kb T=0.2$.\label{fig:analytic}}
\end{figure}

For energy conserving scattering, the model matrix element
Eq.~(\ref{eq:matrix-element-non-momentum-conserving}) only depends on
the transferred energy in the scattering~\cite{note-matrix-element}, since $\Delta
k=\om(1/v_\alpha-1/v_{\bar{\alpha}})$. This
means that the energy integral in the fluctuation spectra
of Eqs.~(\ref{eq:absob-emis-spectra}) can be done analytically upon which it
becomes evident that $\delta f^{(2)}_\alpha(E)\propto
\mathcal{T}(1-\mathcal{T})$. Thus the greater the shot noise of the
QPC, the faster the relaxation is. The elementary scattering
processes leading to relaxation consist of a particle loosing energy in the noisy outer ES and a particle gaining energy in the noiseless inner ES as illustrated on Fig.~\ref{fig:geometry}, (b). The matrix element introduces a
new energy scale $\Delta E\equiv (\h/\ell_p)v_\alpha
v_{\bar{\alpha}}/(v_{\bar{\alpha}}-v_\alpha)$, which limits the
possible amount of energy transferred between the two ES's in
the scattering process since the matrix element is proportional to
$e^{-(\h\om/\Delta E)^2}$. In the limit that $\kb T,|\mu_2-\mu_1|\ll
|\Delta E|$ the distribution functions for the inner and outer ES can be found analytically to be
\begin{align}
\delta f^{(2)}_o(E)=&-\gamma^2\mathcal{T}(1-\mathcal{T})(\mu_2-\mu_1)\nonumber\\
&\times[f_{\mu_2}^0(E)-f_{\mu_1}^0(E)]  \Big[E-\frac{1}{2}(\mu_1+\mu_2)\Big],
\label{eq:df-inter-outer}\\
\delta f^{(2)}_i(E)=&\frac{\gamma^2}{2}\mathcal{T}(1-\mathcal{T})\label{eq:df-inter-inner}\\
\times\bigg\{-[f_{\mu_i}^0&(E)-f_{\mu_i^-}^0(E)]
\big[(\pi\kb T)^2+(E-\mu_i^-)^2\big]\nonumber\\
+[f_{\mu_i^+}^0&(E)-f_{\mu_i}^0(E)]\big[(\pi\kb T)^2+(E-\mu_i^+)^2\big]\bigg\},\nonumber
\end{align}
where $\gamma^2\equiv(2\pi)^2 AL/ [h^4v_\alpha^2v_{\bar{\alpha}}^2]$ 
and $\mu_i^\pm=\mu_i\pm(\mu_2-\mu_1)$ is the maximal and minimal
energy of particles affected by the scattering process in the inner ES.
Here it is seen that the maximal available energy (apart from thermal
excitations of order $\kb T$) is given by the energy difference
$\mu_2-\mu_1$ creating the step distribution. The scattering processes
create a linear slope on the plateau of the distribution of the noisy
outer ES as shown in Fig.\ref{fig:analytic}. The slope crosses the middle of the plateau and it
is proportional to the noise of the QPC and the energy available
$\mu_2-\mu_1$. The inner noiseless distribution gets a tail on both
sides of the Fermi level, which extends over the length of the plateau
$\mu_2-\mu_1$. In the general case, the distribution functions can be
found numerically and the matrix elements in
Eq.~(\ref{eq:matrix-element-non-momentum-conserving}) have to be
included in the calculation, but the transferred energy is still limited by $\Delta E$.  
%






The above perturbative results apply for a short distance $L$ after the
QPC and express the distribution functions at $L$ in terms of the
(unperturbed) distribution functions at the origin. Once the distribution functions
at $L$ are known we can use them to calculate the
distribution functions at a distance $2L$ via Eq.~(\ref{eq:df-inter-second-order}). By iterating this procedure
we can thus describe the effective length dependence of the energy
relaxation. A convenient quantity with which to characterize the
relaxation of 
$f_{\alpha}$ 
at temperature $T$ is given
by the {\em excess temperature} $T_{exc,\alpha}$~\cite{Altimiras:09a} defined as
\begin{equation}
\kb T_{exc,\alpha}\equiv\sqrt{\frac{6}{\pi^2}\int \dd E\Delta f_{\alpha}(E)(E-\tilde\mu_{\alpha})-(\kb T)^2}\,.
\end{equation}
Here $\Delta f_{\alpha}(E)=f_{\alpha}(E)-\theta(\tilde\mu_{\alpha}-E)$ is the difference between the
actual distribution function and a zero temperature Fermi distribution
with the same number of particles and hence
$\tilde\mu_{\alpha}=E_0+\int_{E_0}^{\infty}\dd E f_{\alpha}(E)$, where $E_0$
is chosen such that $f_{\alpha}(E)=1$ for $E<E_0$. $\kb T_{exc,\alpha}$ gives the energy
of the non-thermal excitations in $f_{\alpha}$. The initial excess
temperature right after the QPC of the inner ES
is zero and the one of the outer ES is given by
$\kb T_{exc,o}^0=\{\frac{3}{\pi^2}\mathcal{T}(1-\mathcal{T})\}^{1/2}|\mu_2-\mu_1|$. Because of
    energy conservation, $\sum_\alpha T_{exc,\alpha}$ is a conserved
    quantity in the equilibration process. Furthermore due to
    entropy maximization the excess energy is distributed equally among
    the two ES's, which in the limit of long distances thus
    converge towards Fermi distributions with equal excess
    temperatures given by $\kb T_{exc}^{\infty(2)}=\{\frac{3}{2\pi^2}\mathcal{T}(1-\mathcal{T})\}^{1/2}|\mu_2-\mu_1|$. The excess temperature of the
outer ES measured in~\cite{Sueur:09} does indeed saturate at
large distances toward a finite value. This value is
however found to be systematically lower than the above prediction, for
large voltage biases. Surprisingly
it agrees well with the value $\kb T_{exc}^{\infty(3)}=\{\mathcal{T}(1-\mathcal{T})\}^{1/2}|\mu_2-\mu_1|/\pi$ expected from energy equipartition among
{\em three} instead of only two channels. What could provide the
additional relaxation channel? 
In~\cite{Sueur:09} it has been observed that if the inner ES is forced
to form a short enough closed loop, such that the energy level spacing
of its (discrete) spectrum is larger than the available energy
provided by the voltage bias $\mu_2-\mu_1$, then relaxation of the
outer ES is strongly suppressed. Motivated by this observation, we
suggest that there exist excitations in the bulk~\cite{Granger:09}, which are coupled via long range Coulomb interaction
to both the inner ES and an ES on the opposite side of the sample. As long as the bulk excitations can be created, such a mechanism would allow extra energy to be carried away from the outer ES.

As a first approach we model this extra degree of freedom as an additional
ES coupled to the inner ES only, initially in equilibrium at the electronic
temperature, which we take to be $T=30\,{\rm m K}$. This then contributes an extra collision term in
Eq.~(\ref{eq:df-inter-second-order}) and allows a quantitative
comparison with the experiment~\cite{Sueur:09}. The fitting procedure is
detailed in Supplementary Material and the result is shown in
Fig.~\ref{fig:Texcess-vs-dV}. The best fit
is obtained when the coupling strength to the bulk
excitations is about three times larger than the inter-ES coupling
strength and when
$\Delta E=14.3\,{\rm \mu eV}$, which for $v_o$ and $v_i$ between
$10^4$ and $10^5\,{\rm m/s}$ leads to $\ell_p\geq 0.5\,{\rm \mu m}$. For intermediate
distances (i.e. $2.2\,{\rm \mu m}$ and $4\,{\rm \mu m}$) both the data
and our numerics display a similar
weakly non-linear behavior of the excess temperature as a function of
the voltage bias. 
\begin{figure}[h]
\includegraphics[width=0.45\textwidth]{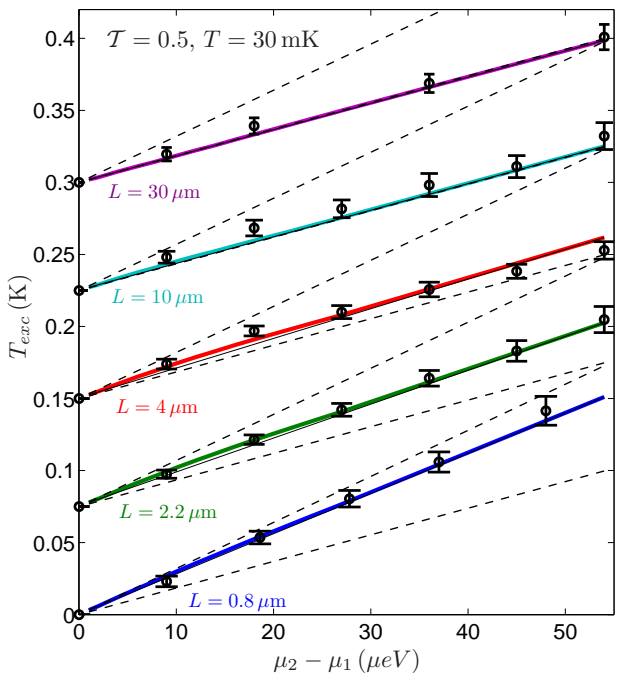}
\caption{(color online). The excess temperature of the outer ES versus the voltage difference across the
QPC. For clarity curves for different voltages have been shifted
upward by $75\,{\rm mK}$. The dashed (black) lines indicate the
initial value $T^{0}_{exc,o}$ (upper lines) and the asymptotic value
$T^{\infty(3)}_{exc}$ (lower lines) as expected from energy equipartition. The thin (black)
lines are guides for the eyes to better see the weak non-linearity
present at intermediate distances (especially for $2.2\,{\rm \mu m}$ and
$4\,{\rm \mu m}$). Experimental data (circles with error bars), courtesy of
F. Pierre et al..\label{fig:Texcess-vs-dV}}
\end{figure}

To conclude, we have computed the distribution functions of two
Coulomb-coupled ES's in the integer
quantum Hall regime. We derived an analytic expression for the leading order correction to the
non-interacting theory, which is present in a system without
translation invariance. We have shown further that the result obtained
in the long distance limit, by iterating the
perturbative solution numerically, is in quantitative agreement with a
recently performed experiment if we take into account the electric
coupling between the innermost ES and bulk excitations. Finally we note that 
measurements of the distribution functions
of both channels for even shorter distances than
$0.8\,{\rm \mu m}$ should allow for further critical testing of our theory.

The authors would like to thank H. Le Sueur, C. Altimiras and
F. Pierre for fruitful
discussion and for sharing their data. This work was supported by the
Swiss NSF.

{\it Note added.---}During completion of this work we became aware of related work
based on plasmon scattering by P. Degiovanni et al.~\cite{Degiovanni:09}.

\bibliography{Bibliography}

\begin{thebibliography}{20}
\bibitem[{not()}]{note-matrix-element}
\bibinfo{note}{This is a common feature of most models of the matrix elements.}
\end{thebibliography}
\bibliographystyle{apsrev}

\cleardoublepage
\newpage

\begin{widetext}

\begin{center}{\begin{singlespace}\Large\bf Supplementary Material for\\ ``Interaction induced edge channel equilibration'' \end{singlespace}}\end{center}
\begin{center}\begin{minipage}[t]{0.9\textwidth}
\hspace{0.2cm}In this Supplementary Material we present details of the
derivation of the leading order correction to the distribution
functions as given in Eq.~(2) of the main text. We also describe the
model used for the Coulomb matrix elements given in Eq.~(4) of the
main text. Finally, we explain the iteration procedure and how we
fitted our theoretical predictions with the experiment
of~\cite{Sueur:09}.
\end{minipage}
\end{center}

\section{The equation of motion approach and the distribution function}
Our starting point is the Hamiltonian describing the
dynamics of the electrons in the inner and outer ES's after the QPC:
\begin{equation}\label{eq:1}
H =
\sum_{\alpha=i,o}\int \dd x\,\psi_{\alpha}^{\dagger}(x)T^{\alpha}(x)\psi_{\alpha}(x)+\frac{1}{2}\sum_{\alpha\beta}\int
\dd x\, \dd x'\,\psi_{\alpha}^{\dagger}(x)\psi_{\beta}^{\dagger}(x')V_{\alpha\beta}(x,x')\psi_{\beta}(x')\psi_{\alpha}(x)\,.
\end{equation}
The first term describes the kinetic energy plus the single particle
potential and the second term the {\em inter} and {\em intra} ES
Coulomb interaction of the outer ($o$) and inner ($i$) ES's. The
intra-ES interaction typically leads only to a small contribution to
the relaxation due to the presence of both the direct and the
exchange term as we shall see shortly. Therefore the focus in the
paper is on the inter-ES interaction $V_{\alpha\bar{\alpha}}$,
using the shorthand notation
$\bar\alpha=\delta_{i\alpha}o+\delta_{o\alpha}i$ (for the opposite
ES of $\alpha$). Introducing the scattering state representation
\begin{equation}
a_{\alpha E}=\int \dd x\,\chi^*_{\alpha
  E}(x)\psi_{\alpha}(x)\quad\Leftrightarrow\quad\psi_{\alpha}(x)=\int\dd E\,
\chi_{\alpha E}(x)a_{\alpha
  }(E)\,,
\end{equation}
makes the single-particle part of the Hamiltonian diagonal. In this
representation, the equation of motion for the annihilation
operators $i\h\p_ta_{\alpha E}(t)=[a_{\alpha E}(t),H]$ in the
Heisenberg picture (i.e.~$A(t)\equiv e^{iHt}Ae^{-iHt}$) becomes
\begin{equation}
i\h\frac{\dd}{\dd t}a_{\alpha E}(t)=Ea_{\alpha E}(t)+\int \dd E'\,
U_{\alpha}(E,E',t)a_{\alpha E'}(t)\,.
\end{equation}
In the case of inter-ES interactions only, the potential operator is
given by
\begin{equation}\label{eq:7}
U_{\alpha}(E,E',t)=\int \dd E_2\dd E_{2'}
V^{\alpha\bar\alpha}_{EE_{2'},E'E_2}a^{\dagger}_{\bar\alpha
  E_{2'}}(t)a_{\bar\alpha E_2^{}}(t)\,,
\end{equation}
where $V^{\alpha\bar\alpha}_{EE_{2'},E'E_2}$, defined in Eq.~(\ref{eq:1}) below, is the Coulomb matrix
element for a transition from an energy $E'$ to $E$ in ES $\alpha$ and
a simultaneous transition from energy $E_2$ to $E_{2'}$ in ES
$\bar\alpha$. A standard perturbation treatment to second order in
$V^{\alpha\bar\alpha}$ leads to the result
\begin{equation}
\delta(E-E')f_{\alpha}^{(2)}(E)\equiv\braket{a_{\alpha
    E}^{\dagger}a_{\alpha E'}^{}}^{(2)}=\delta(E-E')\left(f_{\alpha}^0(E)+\delta f_{\alpha}^{(2)}(E)\right)\,,
\end{equation}
with the inter-ES relaxation given by
\begin{align}\label{eq:8}
\delta f_{\alpha}^{(2)}(E)&=(2\pi)^2\h\int \dd\omega\,\dd
E'|V^{\alpha\bar\alpha}_{EE'+\h\omega,E+\h\omega
  E'}|^2\\
&\times \Big[f_{\alpha}^0(E+\h\omega)[1-f_{\alpha}^0(E)]f_{\bar\alpha}^0(E')[1-f_{\bar\alpha}^0(E'+\h\omega)]-f_{\alpha}^0(E)[1-f_{\alpha}^0(E+\h\omega)]f_{\bar\alpha}^0(E'+\h\omega)[1-f_{\bar\alpha}^0(E')]\Big]\,.\nonumber
\end{align}
The combination of Fermi functions which appears here ensures the
Pauli exclusion principle. Furthermore one can easily show that
\begin{equation}
\braket{\delta U_{\alpha}(E,E',\omega)^{(1)}\delta
  U_{\alpha}(E',E,\omega')^{(1)}}=(2\pi)^2\h\delta(\omega+\omega')\int
\dd E_{2'}|V^{\alpha\bar\alpha}_{E'E_{2'}+\h\omega,EE_{2'}}|^2f_{\bar\alpha}^0(E_{2'})[1-f_{\bar\alpha}^0(E_{2'}+\h\omega)]\,,
\end{equation}
where $\delta
U_{\alpha}(E,E',\omega)^{(1)}=U_{\alpha}(E,E',\omega)^{(1)}-\braket{U_{\alpha}(E,E',\omega)^{(1)}}$
and $U_{\alpha}(E,E',\omega)^{(1)}$ is the Fourier transform of the first order expansion of~(\ref{eq:7}). This then
leads immediately to Eq.~(2) of the main text.

The relaxation due to intra-ES interactions can be found in the same
way. The result is as in Eq.~(\ref{eq:8}) with the important
replacement:
\begin{align}
|V^{\alpha\bar\alpha}_{EE'+\h\omega,E+\h\omega E'}|^2 \rightarrow
\frac{1}{2}
|\underbrace{V^{\alpha\alpha}_{EE'+\h\omega,E+\h\omega
E'}}_{\textrm{direct term}}
-\underbrace{V^{\alpha\alpha}_{E'+\h\omega E,E+\h\omega
E'}}_{\textrm{exchange term}}|^2,
\end{align}
where the two final states were exchanged in the second term called
the exchange term. The appearance of both the direct and the
exchange interaction is due to the fact that the electrons are in
the same conductor and hence quantum mechanical exchange processes
are important. For the simple matrix element model used below, the
direct and exchange elements exactly cancel. Although this
cancellation is model dependent, the intra-ES contribution is strongly
suppressed as a general feature and is therefore safely
neglected in the present work.

Finally, we note that the generalization to more than two coupled
ES's is straightforward and formally results in additional collision
terms in Eq.~(\ref{eq:8}). In the following, we make use of this
generalization to incorporate the effect of an extra relaxation
channel.

\section{Model for the Coulomb matrix elements}
The interaction matrix element is
\begin{equation}\label{eq:1}
V^{\alpha_1\alpha_2}_{E_{1'}E_{2'},E_1E_2}\equiv\int \dd x\,\dd
x'\chi^*_{\alpha_1^{}E_{1'}}(x) \chi^*_{\alpha_2^{}E_{2'}}(x')
V_{\alpha_1\alpha_2}(x,x')
\chi^{}_{\alpha_2^{}E_2^{}}(x')\chi^{}_{\alpha_1^{}E_1^{}}(x),
\end{equation}
where $V_{\alpha_1^{}\alpha_2^{}}(x,x')$ is the effective Coulomb
interaction between an electron at $x$ in ES $\alpha_1$ and an
electron at $x'$ in ES $\alpha_2$. In the adiabatic edge channel
description~\cite{Beenakker:91}, ES's follow the equipotential lines
in the sample. Due to impurities near the edge and to the variation
of the confining potential, the inter-ES distance is not constant in
the direction of propagation. In other words, translation invariance
is broken
and momentum is not conserved in the collision process.
For the sake of generality, we do not model a specific geometry nor
impurity distribution, but instead simply average over the geometry.
%
This 
naturally introduces a momentum breaking correlation length scale
$\ell_{p}$ quantifying the amount of momentum breaking present in
the system.
\begin{center}
\begin{SCfigure}[1][h]
\includegraphics[width=0.3\textwidth]{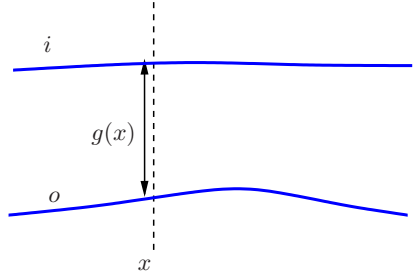}\caption{(color
  online). Schematics
  of two co-propagating edge states with varying inter-edge state distance.\label{fig:a1}}
\end{SCfigure}
\end{center}
As a specific simple example,
we consider the interaction between points with the same $x$
coordinate as illustrated in Fig.~\ref{fig:a1},
i.e.~$V_{\alpha\bar\alpha}(x,x')=\delta(x-x')g(x)\,$, where $g(x)$
is an unknown function of the interaction strength variation along
the ES's. This approximation is justified for very good screening.
However, it is important to emphasize~\cite{Pothier:97} that this simplification is
not of vital importance for the physics discussed here, i.e.~if we
consider the poor screening limit, then the same physical effect
of the broken translation invariance remains as can be shown from
a more complete calculation.
We assume linear dispersion relations with constant velocities
$v_{\alpha}$ and use a single particle basis of plane wave states
\begin{equation}\label{eq:3}
  \chi_{\alpha E}(x)=\frac{1}{\sqrt{h v_{\alpha}}}e^{ik_{\alpha}(E)x}\,,\quad\text{with}\quad
  k_{\alpha}(E)=\frac{E}{\h v_{\alpha}}\,.
\end{equation}
Note that these states are normalized to have equal current and obey
$\int\dd x \chi^*_{\alpha E}(x)\chi_{\alpha E'}(x) =\delta(E-E')$.
With the above assumptions, we obtain
\begin{equation}\label{eq:4}
V^{\alpha\bar\alpha}_{E_{1'}E_{2'},E_1E_2}=\frac{1}{h^2v_{\alpha}v_{\bar\alpha}}\int_{-L/2}^{L/2}
\dd x\,e^{i\Delta k x}g(x)\,,
\end{equation}
where the amount of broken momentum conservation $\Delta
k=k_\alpha(E_1)+k_{\bar\alpha}(E_2)-k_\alpha(E_{1'})-k_{\bar\alpha}(E_{2'})$
is introduced and $x$ is integrated over the region of relaxation of
length $L$. Next the geometrical averaging of the
squared matrix element
$|V^{\alpha\bar\alpha}_{E_{1'}E_{2'},E_1E_2}|^2$ is performed by assuming that
the deviation $\Delta g(x)=g(x)-g_0$ is Gaussian distributed, i.e.
$\overline{\Delta g(x)\Delta
g(x')}=A/(\sqrt{2\pi}\ell_{p})\exp[-(x-x')^2/(2\ell_{p}^2)]$, as
described in the main text.
This gives two contributions to
$\overline{|V^{\alpha\bar\alpha}_{E_{1'}E_{2'},E_1E_2}|^2}$: one with $\Delta
k=0$ and one where $\Delta k\neq 0$ is possible. The latter is given
by Eq.~(4) of the main text. Note that it is proportional to $L$, because interaction is included only over a region of
length $L$.

In this work, linear dispersions with different velocities are used,
which leaves no phase space for scattering in the momentum
conserving case $\Delta k=0$ due to simultaneous energy and momentum
conservation. In general, the dispersions are not linear and this
changes the phase space constraints due to momentum
conservation in the translation invariant limit. For instance,
for a spin-split quadratic dispersion there would only be a single
possible final state, where the two initial momenta are simply
interchanged. This would lead to a very small resonant-like feature on
the plateau of the outer ES distribution function, which could be
tuned along the plateau by the chemical potential of the inner ES
$\mu_i$. Any dispersion with a positive curvature will lead to the same
result. However, this is a very small increase of the phase space compared to
the increase introduced by the non-translation invariant physics
and importantly, this is \emph{not} what is observed experimentally.
Therefore for the non-translation invariant physics the curvature
of the dispersion is without importance and for simplicity the
dispersion is linearized.



\section{Numerical iteration procedure}
Eqs.~(2) and~(3a-b) of the main text express the variation of the distribution
functions of the ES's after a short distance $L$ as a
functional of the distribution functions at the origin. We thus have a system of coupled
equations for the approximate distribution functions after a distance
$L$, which is valid to second order in the interaction and has the form
\begin{equation}
f_{\alpha,L}(E) = F[f_{\alpha,0},f_{\bar\alpha,0}]+O(V^3)\,.
\end{equation}
Iterating this, we obtain a recursive relation
\begin{equation}
f_{\alpha,nL}(E) \approx F[f_{\alpha,(n-1)L},f_{\bar\alpha,(n-1)L}]\,,
\end{equation}
where $n$ is the iteration number. Thus we can calculate an
approximation to the distribution functions at distances~$nL$.

A typical
result of this iterative procedure for three coupled ES's is shown in Fig.~\ref{fig:a3}, where we show the evolution of the excess
temperature of the inner ES, the outer ES and the additional
relaxation channel (denoted with ``bulk'') as a
function of the iteration number. For the numerical integration, we
use a three points Simpson extrapolation rule. We see that the inner
ES and the additional relaxation channel, which are initially at
equilibrium both gain energy since their excess temperatures
increase. This increase happens at the expense of the initially out of
equilibrium outer ES, which loses energy during the relaxation
processes and thereby its excess temperature decreases. All three converge
towards the temperature $k_BT_{exc}^{\infty,(3)}= \{\mathcal{T}(1-\mathcal{T})\}^{1/2}|\mu_1-\mu_2|/\pi$, expected from energy
  equi-partition. Inset (a) shows the distribution
functions after $1$ iteration and inset (b) after $54$ iterations. Note
in particular that all three distribution functions are found to converge
towards Fermi functions (dashed (black) curves in inset (b) which are indistinguishable from the calculated distributions). While
the chemical potentials of the equilibrated distribution functions of
the inner ES and of the additional relaxation channel remain constant, the chemical
potential of the equilibrated outer ES is given by
$\mu^{\infty}_o=\mu_1+\mathcal{T}(\mu_2-\mu_1)$, in agreement with
particle conservation within each ES and thereby current
conservation. Finally, note that because the velocity is independent
of energy, the relaxation process is independent
of the chemical potentials of the
inner ES and of the additional relaxation channel.
\begin{center}
\begin{figure}[h]
\includegraphics[width=0.7\textwidth]{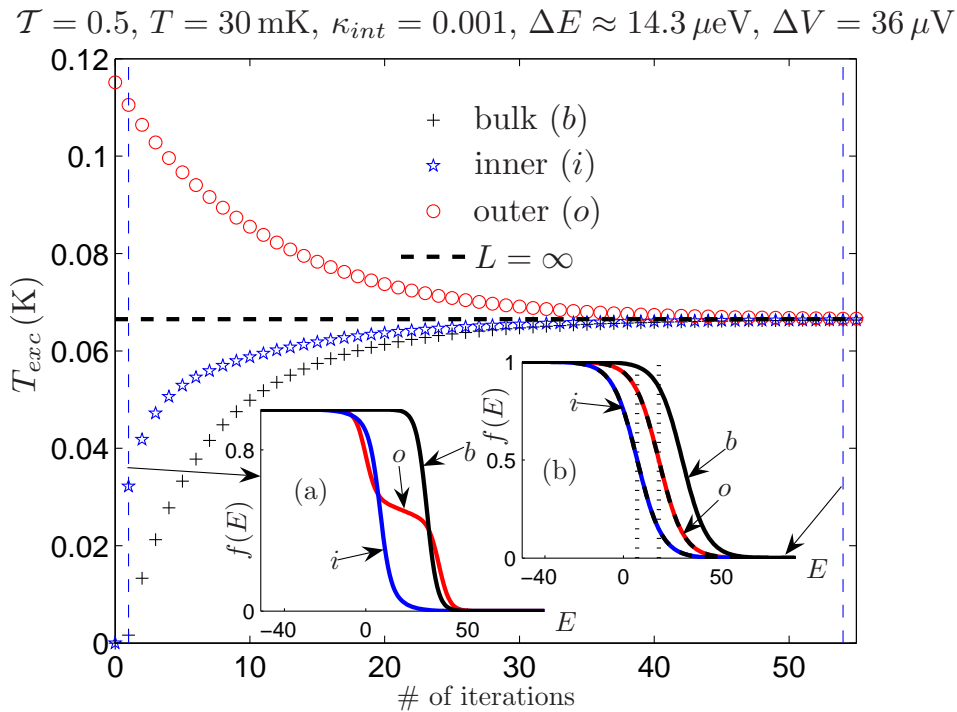}\caption{(color
  online). Excess
  temperature as a function of the iteration number. Inset (a)
  shows the distribution functions after $1$ iteration (i.e. the
  second order perturbation theory result) and inset (b) shows the
  distribution functions after $54$ iterations where the edge states and the additional relaxation channel have relaxed into hot Fermi
  functions with the same temperature. The dashed
  vertical lines in inset (b) give the asymptotic values of the chemical potential
  of the two edge states ($\mu_i^{\infty}=\mu_i$ and $\mu_{o}^{\infty}$).\label{fig:a3}}
\end{figure}
\end{center}

\section{Fitting the theory to the experiment}

A direct comparison of our theory with the
experiment~\cite{Sueur:09} is difficult because the absolute
values of the interaction matrix elements
among the ES's and between
the inner ES and the bulk are unknown. Furthermore, we lack the knowledge
of the proportionality constant $\alpha$ between the distance and the number of
iterations. Nevertheless, using the matrix elements and the
proportionality factor $\alpha$ as free parameters, we could fit all
of the data for different voltage biases and lengths
simultaneously, i.e. using the same parameters to fit different values of $\mu_2-\mu_1$ and $L$. We find the best fit is obtained when the interaction strengths
between the two outer ES's is about three times weaker than the interaction strength between
the inner ES and the bulk excitations. The optimal value for the
allowed energy transfer is found to be $\Delta E\approx 14.3\,{\rm \mu
  eV}$. Following the literature~\cite{McClure:09}, and taking the ES velocities to be between $10^4$ and $10^5\,{\rm
  m/s}$ this leads to a lower bound of $0.5\,{\rm \mu m}$ for the
momentum breaking length $\ell_p$. This lower bound is smaller than
the shortest measured propagation
length of $0.8\,{\rm \mu m}$ and roughly one hundred times larger than the estimated magnetic
length.

The length dependence of the excess temperature for the optimal
parameters and for different values of $\Delta V=\mu_2-\mu_1$,
is shown in Fig.~\ref{fig:a2}, where it is compared with the
experiment. The thin dashed (red) curves give a least square exponential fit to the
data. To obtain the excess temperature at an
arbitrary distance $x$, we linearly interpolate the excess temperature from the numerical values
obtained from the iteration procedure at distances $L\lfloor
x/L\rfloor$ and $L(\lfloor
x/L\rfloor+1)$. This is how we obtain Fig.~(3) of the main text.

Finally, in Fig.~\ref{fig:4} we compare the cases with and without the extra
relaxation channel, using the same optimal parameters as before. At small
voltage biases, the data agrees well with the case where relaxation
occurs only between the inner and the outer ES's, while for larger voltage
biases it agrees at long distances only with the case where all three relaxation channels are
effective. This clearly indicates that the energy is split among three and not two systems in the fully equilibrated limit.
\begin{center}
\begin{SCfigure}[0.8][h]
\includegraphics[width=0.5\textwidth]{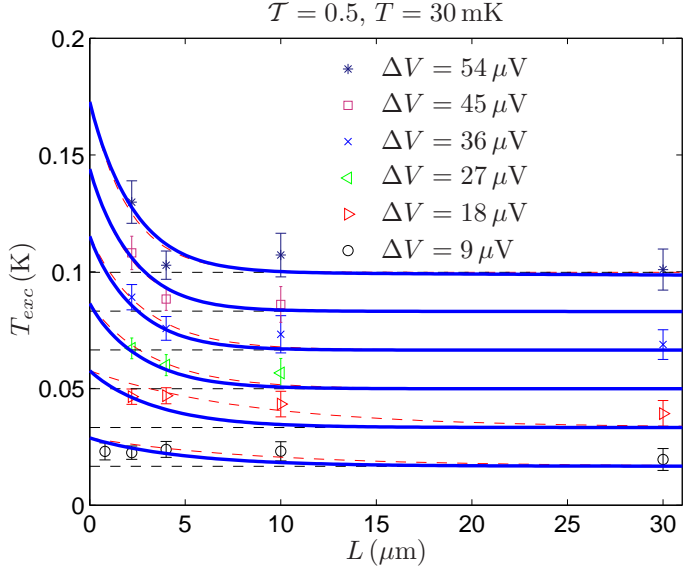}\caption{(color
  online). Length dependence of the excess
  temperature of the outer edge state. Comparison
  between theory and experiment. Symbols with
  errorbars show the measurement results of~\cite{Sueur:09}. The thin dashed (red)
  curves show a least square exponential fit to the data.\label{fig:a2}}
\end{SCfigure}

\begin{SCfigure}[0.8][h]
\includegraphics[width=0.5\textwidth]{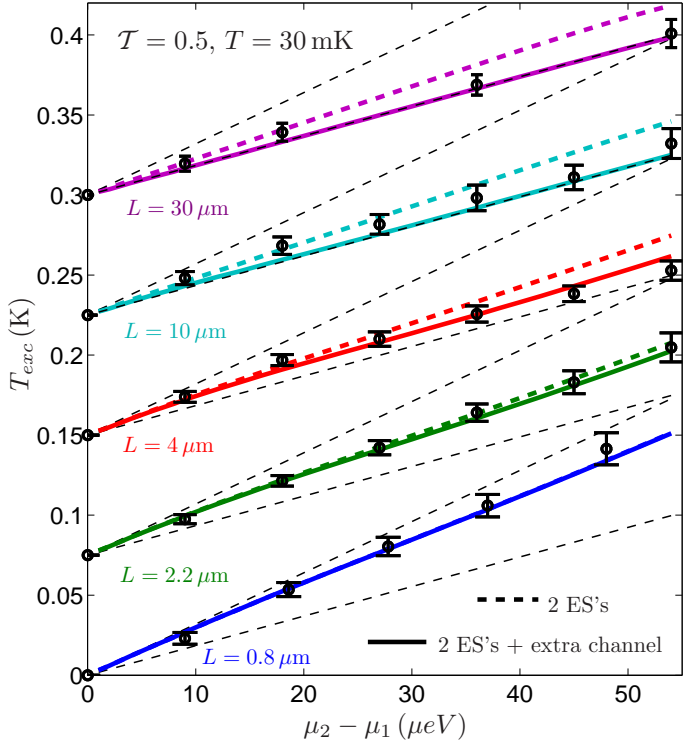}\caption{(color
  online). Comparison
between the models with (full thick curves) and without (dashed thick
curves) additional relaxation channel. For small voltage biases
$\mu_2-\mu_1$ the data of~\cite{Sueur:09} (black symbols with errorbars) is in good agreement with the model
including relaxation among the inner and outer edge states only, while for
larger voltages and long distances the agreement is markedly better
with the model including an extra relaxation channel.\label{fig:4}}
\end{SCfigure}
\end{center}

\end{widetext}

\end{document}